\documentclass[aps,prl,twocolumn,groupedaddress]{revtex4}

\usepackage{graphicx}
\usepackage{dcolumn}
\usepackage{bm}

\begin{document}

\title{\large Observation of McMillan-Rowell like
oscillations in underdoped YBCO junctions oriented along the node
of the d-wave order parameter}

\author{L. Shkedy, P. Aronov, G. Koren and E. Polturak}
 \affiliation{Physics Department, Technion - Israel Institute of Technology Haifa,
 32000, ISRAEL}

 \email{gkoren@physics.technion.ac.il}
 \homepage{http://physics,technion.ac.il/~gkoren}

\date{\today}

\begin{abstract}
Dynamic resistance spectra of ramp junctions made of underdoped
$YBa_{2}Cu_{3}O_y$ electrodes and Ga-doped YBCO barrier are
reported. Series of equidistant peaks were observed in these
spectra in junctions oriented along the node direction. Junctions
with different barrier thickness $d_N$, showed that the distance
between adjacent peaks scales inversely with $d_N$. The peaks were
thus identified as due to McMillan-Rowell like oscillations in the
barrier. Analysis of the series of peaks yields an upper limit of
about 3.7 meV on the value of the energy gap along the node. We
attribute this small gap to the $is$ component of the order
parameter of underdoped YBCO near the interface of the junctions.

\end{abstract}

\maketitle{\large   INTRODUCTION}\\

The fine details of the symmetry of the order parameter in the
high temperature superconductors (HTS) are still under debate.
Determining the exact symmetry could be important for
understanding the mechanism of the high $T_c$ superconductors.
Several experiments show that the dominant component of the order
parameter in the HTS materials has a $d_{x^2-y^2}$-wave symmetry,
as summarized in a review article by Tsuei and
Kirtley.\cite{Tsuei} Other experiments are consistent with the
existence of an additional sub-dominant component on the surface
of the HTS, of $is$ or $id_{xy}$
nature.\cite{Covington,Krupke,Sharoni,Koren2002,KorenJLTP}
Tunneling measurements of underdoped junctions show that in
addition to the $d_{x^2-y^2}+is$ gap, a large gap which can be
attributed to the pseudogap is also
present.\cite{DeutscherNature,Krasnov,KorenPseudogap} The
magnitude of the $is$ component in these
studies,\cite{Covington,Krupke,Sharoni,Koren2002,KorenJLTP} was
found to be in the range of ~1-3 meV as determined from the peak
to peak distance in the conductance curves. Recent self-consistent
calculations using Bogoliubov-DeGennes type equations, led to a
good fit of the data assuming a pure d-wave symmetry in the bulk,
and coexisting $d_{x^2-y^2}$ and $is$ order parameters near the
interface.\cite{Lubimova} The magnitude of the $is$ gap resulting
from these simulations is 2.6$\pm 0.1$ meV. In the present
experiment we observed series of geometrical resonances in the
dynamic resistance spectra of node junctions. From these series,
we find an upper limit of 3.7 meV on the energy of the
sub-dominant s-wave component of the gap, which is consistent
with the simulations results.\\

\begin{figure}
\includegraphics[height=6cm,width=8.5cm]{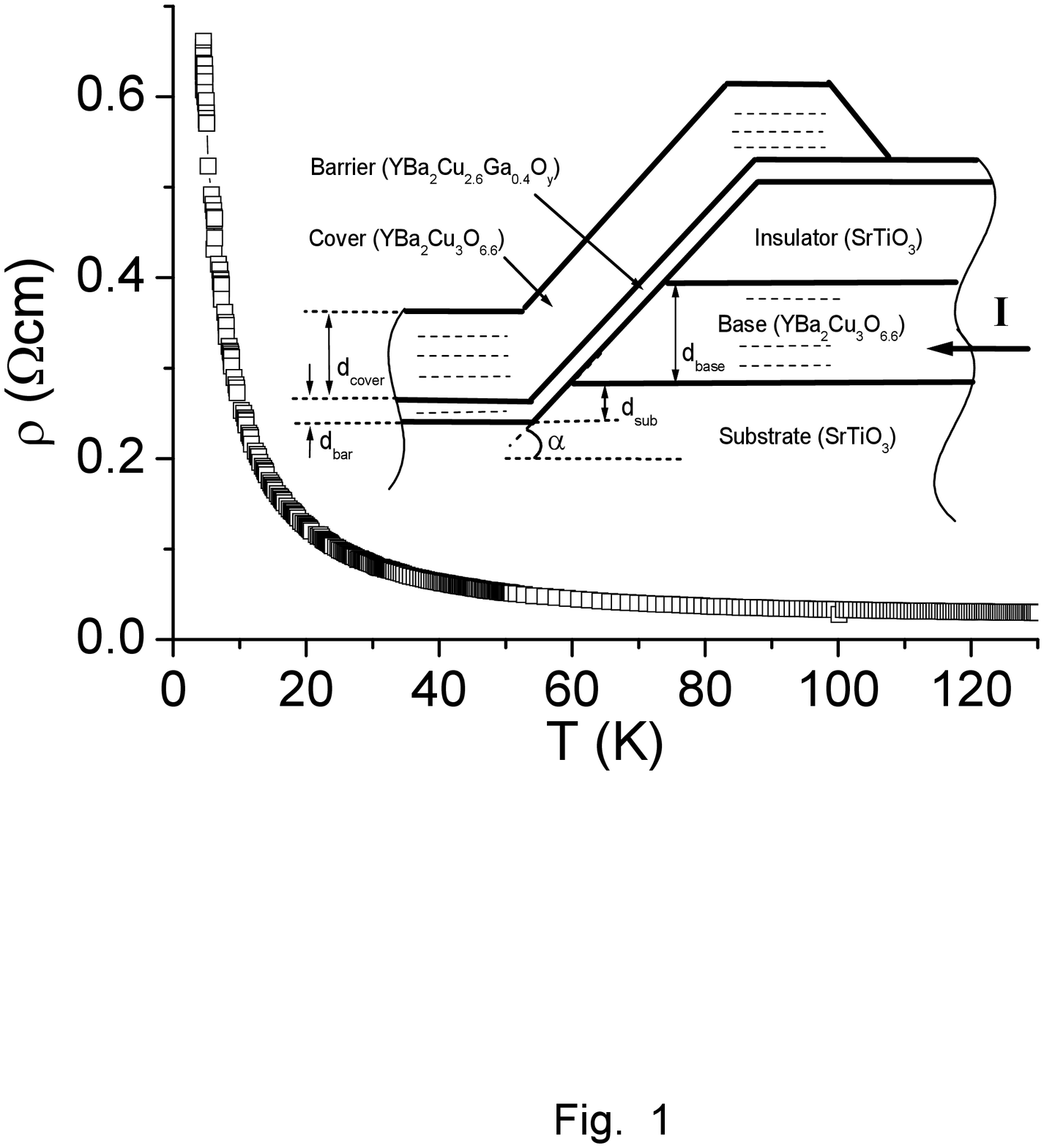}
\caption{\label{fig:epsart} Resistivity versus temperature of a
Ga-doped YBCO film. The resistivity values were obtained by
averaging over six microbridges which were patterned in the ~80 nm
thick film. The inset shows a schematic cross section of a ramp
type junction. The tunneling current flows along the a-b plane,
while c-axis current is suppressed in the junction due to the
thick insulating STO layer.}
\end{figure}

 \maketitle{\large   EXPERIMENTAL}\\

The junctions used in the present study are of the same type of
ramp junctions used
before,\cite{KorenPseudogap,NesherTomasch,Koren2002,KorenJLTP,NesherSubHarmonic}
but instead of the Fe-doped YBCO barrier, we used a Ga-doped YBCO
barrier. This was done in order to check if the appearance of the
s-wave component depends on the nature of the barrier. In
particular, the possible presence of magnetic impurities in the Fe
doped barrier, is probably absent when a Ga doped barrier is used.
Therefore, observation of a signature of an $is$ gap, in both type
of junctions, is indicative that magnetic effects in the barrier
are not likely to be the source of this gap. Fig. 1 shows the
resistivity versus temperature of a blanket
$YBa_2Ga_{0.4}Cu_{2.6}O_y$ film deposited on (100) $SrTiO_3$ (STO)
wafer, annealled {\em in-situ} under the same annealing conditions
as for obtaining YBCO films with $T_c=60K$. This barrier material
behaves like a Mott insulator (MI) with variable range hopping
(VRH) in 3D with lg$(\rho) \propto T^{-1/4}$ for the whole
temperature range. The resistivity value at 2K is of about
$0.6\Omega cm$, which is more than an order of magnitude higher
than that of $YBa_2Fe_{0.45}Cu_{2.55}O_y$ used previously as the
barrier layer.\cite{NesherSubHarmonic} The geometry of the
junction is described schematically in the inset of Fig. 1. All
the YBCO and doped YBCO layers are epitaxial with the $c-axis$
normal to the wafer. The two superconducting YBCO electrodes are
coupled in the
$a-b$ plane via a thin barrier layer of the Ga-doped YBCO film.\\

The multi-step junction preparation process was described
before.\cite{NesherTomasch} Briefly, we first prepared  by laser
ablation deposition the base electrode which was composed of a
bilayer of STO on YBCO on (100) STO wafer. Patterning of the base
electrode was done by photolithography and Ar ion beam milling.
The ramps of the junctions were patterned along the node direction
of the YBCO film. After a thorough cleaning process, the cover
electrode was deposited. This included the barrier layer, a second
YBCO film, and an Au layer on top. The cover electrode was then
patterned to produce the final junction layout, as well as the
four gold pads for each junction. All junctions in the present
study had the same 90nm thick YBCO electrodes (base as well as
cover), the same lateral width of $5\,\mu m$, and varying barrier
thickness. The resistance versus temperature of the junctions was
measured using the standard four probe technique, and the dynamic
resistance was measured using a standard ac modulation
technique.\\

\begin{figure}
\includegraphics[height=6cm,width=8.5cm]{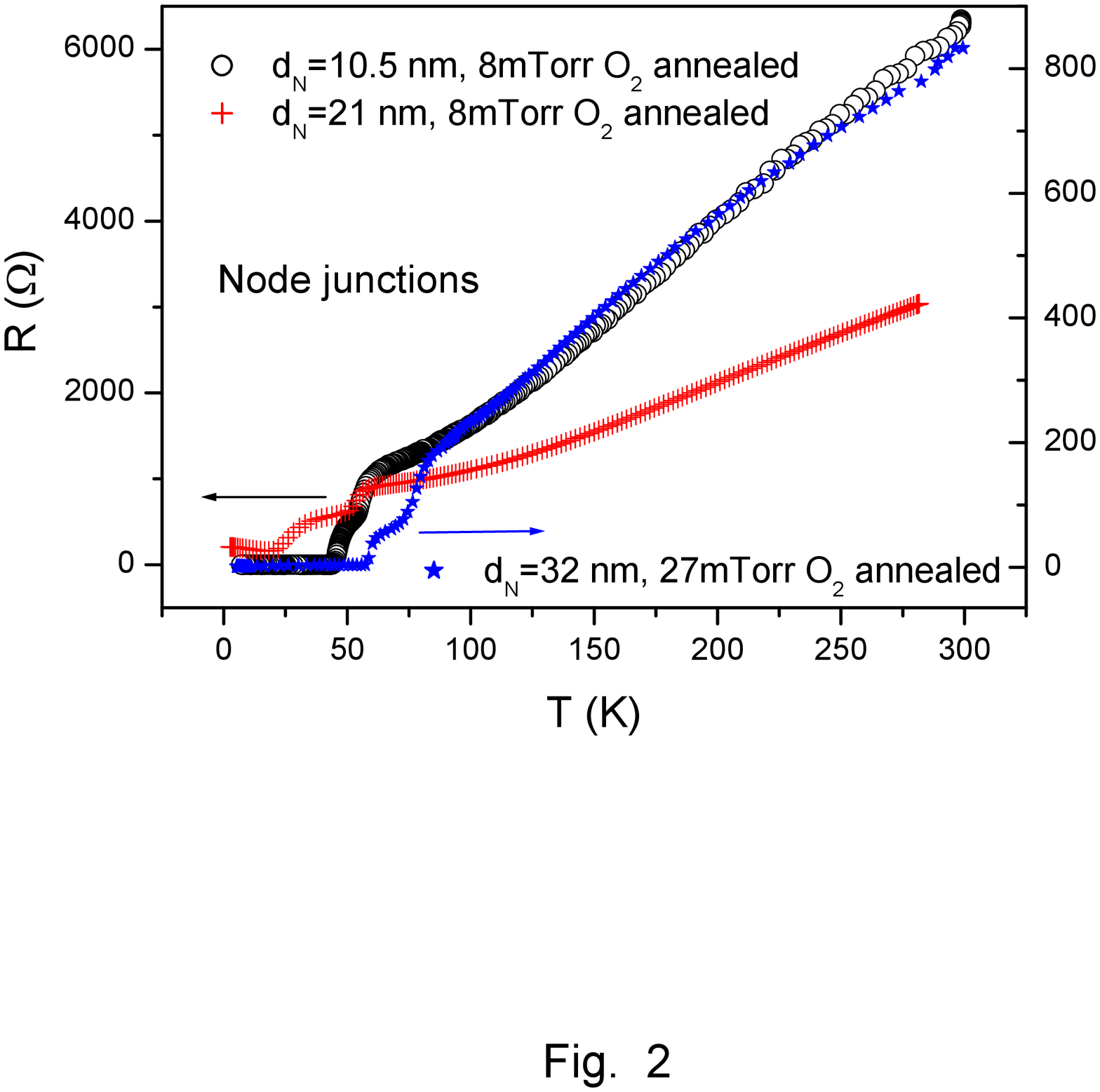}
\caption{\label{fig:epsart} Resistance versus temperature of three
ramp junctions with 10.5, 21 and 32 nm thick Ga-doped YBCO
barrier, annealled in 8, 8 and 27 mTorr of oxygen pressure,
respectively. The three junctions were patterned along the node
direction at an angle of $45^{0}$ to the main axis. The junction
with the 10.5nm thick barrier had longer leads which led to a
higher normal resistance. }
\end{figure}

\maketitle{\large RESULTS AND DISCUSSION}\\

Fig. 2 shows the measured resistance as a function of temperature
of three node junctions with 10.5, 21 and 32 nm thick barriers.
The first two junction were annealled in 8 mTorr of flowing
oxygen, and the last one in 27 mTorr oxygen flow, respectively.
The corresponding normal state resistance of these three junctions
is typical of underdoped $YBa_{2}Cu_{3}O_y$ with $y\sim6.55$ for
the low oxygen annealled ones and $y\sim6.85$ for the richer
oxygen junction.\cite{Segawa} We note that the higher normal state
resistance of the junction with the thinnest barrier is due to its
longer leads (a different photolithographic mask pattern was
used). One can observe two distinct transitions in the resistance
of each junction. In the two oxygen deficient junctions (8 mTorr
annealing), the transition temperature $T_c$ of the electrodes
occurs at approximately 55K, while in the third oxygen rich
junction (27 mTorr annealing) the electrodes become
superconducting already at 80K. The transitions seen at 40 and 25K
in the first two junctions, and at 60K in the third one, result
presumably from an apparent proximity effect in the barrier of the
different junctions. At low temperatures, the junction with the
10.5nm thick barrier shows a critical current of about 0.5mA at 5K
which yields a critical current density of $\sim 1.1\times
10^4\,A/cm^{2}$. The other two junctions with the 21 and 32nm
thick barriers are resistive at low temperatures and have
resistance values at low bias of about 200 and 2$\Omega$,
respectively. In these two cases, the barriers exhibit insulating
behavior with the resistance either increasing slightly with
decreasing temperature or staying almost
constant.\\

\begin{figure}
\includegraphics[height=6cm,width=8.5cm]{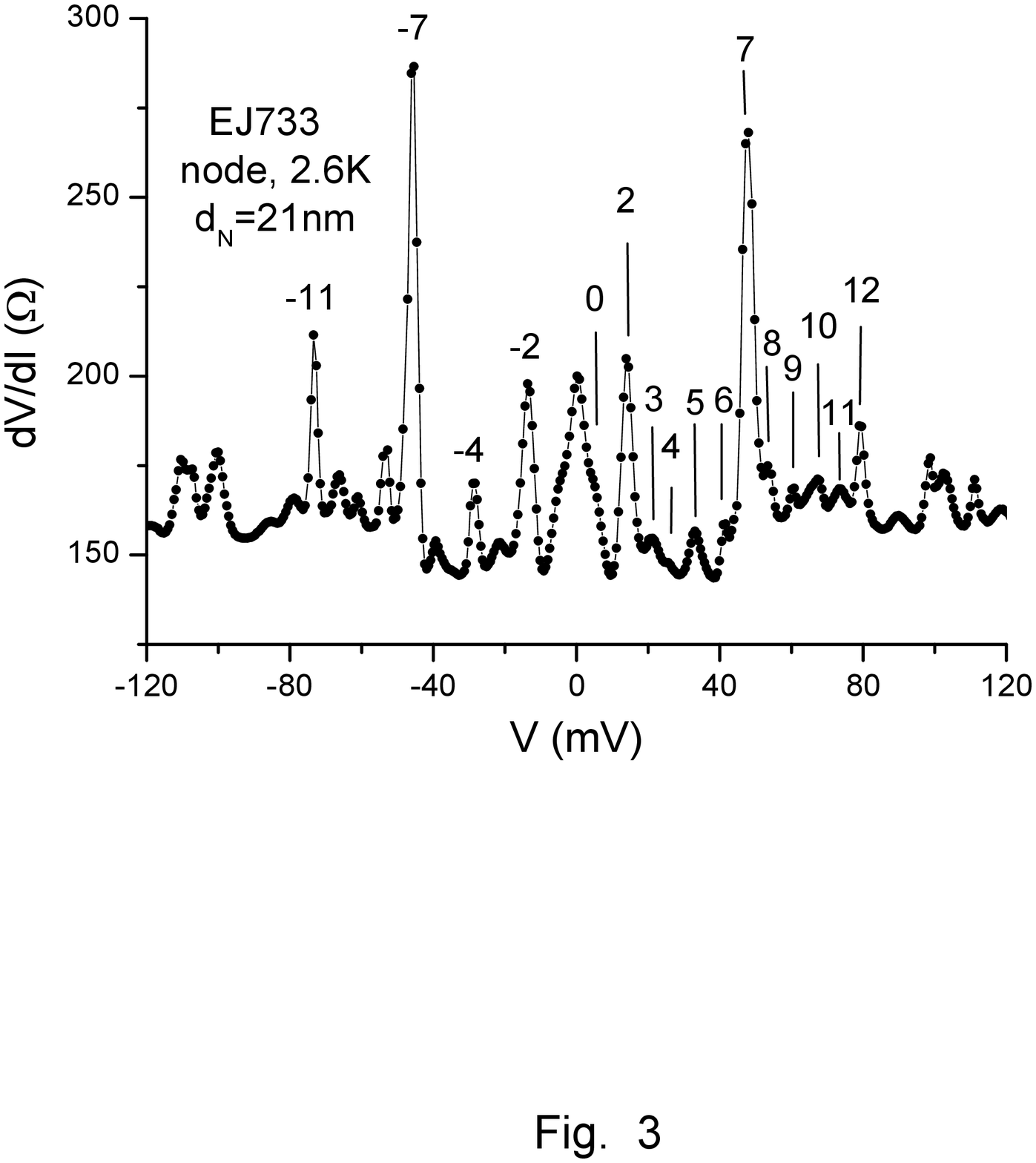}
\caption{\label{fig:epsart} Dynamic resistance spectra of the
junction in Fig. 2 with the 21 nm thick barrier and 8 mTorr oxygen
annealing pressure. }
\end{figure}

\begin{figure}
\includegraphics[height=6cm,width=8.5cm]{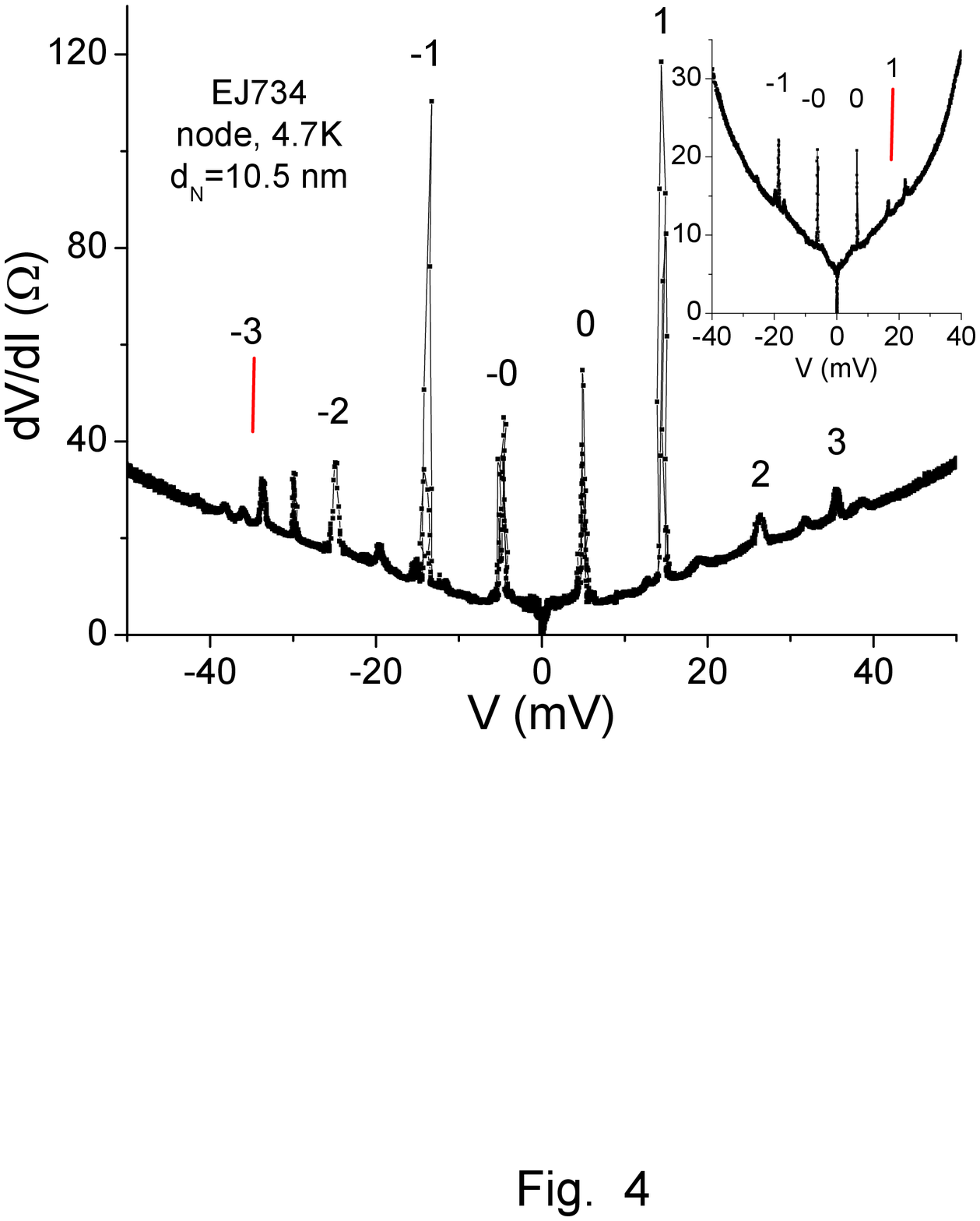}
\caption{\label{fig:epsart} Dynamic resistance spectra of the
junction in Fig. 2 with the 10.5 nm thick barrier and 8 mTorr
oxygen annealing pressure (three traces, main panel, peak no. -3
is missing). In the inset, the results of another node junction on
the same wafer are shown (two traces, peak no. 1 is missing). }
\end{figure}

The dynamic resistance spectra of the two oxygen deficient node
junctions with 21 and 10.5nm thick barriers are shown in Figs. 3
and 4, respectively. The different behavior at low bias, namely
tunneling like in Fig. 3, and critical current and zero bias
conductance peak (ZBCP) in Fig. 4, is due to the much higher
normal resistance of the first junction ($R_N(3K)\sim 200\Omega$).
We prepared another node junction with a 32nm thick barrier, but
annealled it in a higher oxygen pressure of 27mTorr. This was done
in order to avoid a very high $R_N(5K)$, and yielded $R_N(5K)\sim
2\Omega$ (see Fig. 1). The dynamic resistance spectra of this
junction (not shown here) was similar to that of Fig. 3, but with
a ZBCP and a more closely spaced series of peaks. In all three
junctions, the position of the series of peaks on the voltage axis
in the dynamic resistance curves seem to be almost independent of
the oxygen content, or the presence of a ZBCP. It was however
strongly dependent on the thickness of the barriers. In order to
determine the origin of the series of peaks, we plotted the peak
voltages versus peak number for each of the three junctions. It is
generally nontrivial to associate a peak number to each peak
because not all of them are present or have the same intensity.
Some of the peaks are missing, some are enhanced while others can
also overlap and interfere with one another. Nevertheless, by
comparison with data of other node junctions, we could determine
the peak numbers properly, and in Fig. 5 we plot the peak voltages
of four series versus the peak number including the data of Figs.
3 and 4. Fig. 5 shows that each series of peaks appears with a
constant voltage difference between adjacent peaks. Linear fits of
the four series yield $V_n=5.1+10.1n$ and $V_n=6+12n$ for the two
junctions in Fig. 4 with the thinnest barrier, $V_n=1.9+6.5n$ for
the junction in Fig. 3, and $V_n=1.9+4.5n$ for the fourth junction
with the 32nm thick barrier. The intersects at n=0 (5.1, 6, 1.9
and 1.9 mV) are of the same order of magnitude as the bias
voltages of the corresponding first peaks in the series (5, 6, 3.4
and 2.4 mV). Fig. 5 also shows that the adjacent peaks spacing is
sensitive to the barrier thickness. Its ratio in the three types
of junctions is 10.1-12:6.5:4.5 which is approximately equal to
the inverse ratio of the barrier thickness (1/10.5):(1/21):(1/32).
Since the superconducting electrodes in all our junctions have the
same thickness (90nm), the above result indicates that the series
of peaks in the dynamic resistance spectra originate in
geometrical resonances in the barrier layer. Next we discussed
this result in the context of the nature of the present S/MI/S
junctions (MI is Mott insulator with VRH).\\

\begin{figure}
\includegraphics[height=6cm,width=8.5cm]{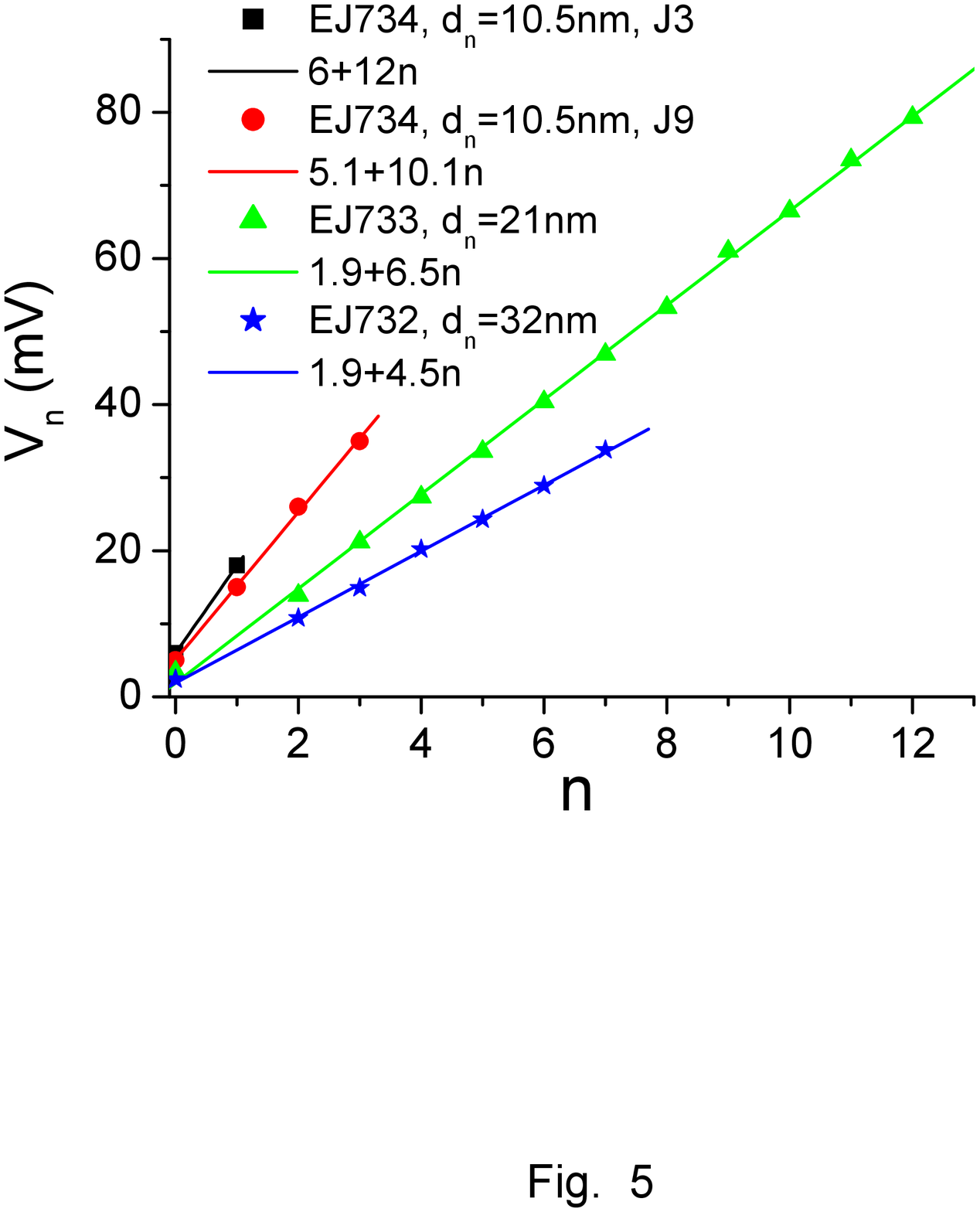}
\caption{\label{fig:epsart} Peak voltages of the series in Figs. 3
and 4 versus the peak number, together with the series of peaks of
the oxygen rich junction in Fig. 2 with the 32 nm thick barrier.
The straight lines are linear fits to the data. }
\end{figure}

It has already been demonstrated in the past that a-axis
YBCO/$PrBa_{2}Cu_{3}O_{7-\delta}$/YBCO junctions which are
basically S/MI/S junctions as we have here, carry significant
critical currents $I_c$ at low temperatures even when the barriers
are up to 100nm thick.\cite{Yoshida} Surprisingly for these kind
of junctions, it was found that $I_c(T)$ behaves as
exp$[-aT^{0.5}]$ where $a$ is a constant, which is exactly the
expected behavior for SNS type junctions with a normal metal
barrier in the dirty limit. There are many more reports on
observations of a long range proximity effect and Andreev
reflections in similar type of junctions with insulating VRH
barriers.\cite{Kabasawa,Adrian,Frydman}  The puzzling question is
why such barrier materials with a resistivity of 0.1-1 $\Omega cm$
which is two to three orders of magnitude higher than $\sim 1
m\Omega cm$ the maximum resistivity a metal can have, when in
contact with a superconductor should behave like normal metals?
This is a complicated problem to deal with theoretically, and only
sketchy reports on this issue exist.\cite{Tanaka} We shall not
attempt to speculate here what is the reason for this normal metal
like behavior of the VRH barriers in the S/MI/S junctions, but
simply take it as given, and use as a first approximation the
formulas derived for SNS junctions. We note however, that unlike
the previous reports where the HTS electrodes of the S/MI/S
junctions were close to optimal
doping,\cite{Yoshida,Kabasawa,Adrian} in our case the underdoped
60K YBCO based junctions yield critical current only for the
thinnest 10.5nm barrier. This can perhaps be related to the fact
that the present node junctions have a weaker superconductivity
coupling and shorter proximity penetration lengths, but detailed
investigation of this effect is outside of the scope of the
present study. We thus proceed with the treatment of our
S/MI/S junctions, along the line used with SNS junctions.\\

We now discuss four possible types of geometrical resonances in
SNS junctions that can lead to similar series of peaks in the
dynamic resistance spectra like we observed here. Two are due to
sub-gap scattering processes of quasiparticles, and two to above
gap processes. The first two involve sub-harmonic resonances,
which are caused by multiple Andreev reflections, or
DeGennes-Saint James bound
states.\cite{NesherSubHarmonic,Andreev,DGSJboundStates} Both
result in series of peaks which are not equally spaced, and
therefore are not similar to the presently observed series.
Moreover, the junctions in the present study are oriented along
the node direction where the d-gap vanishes, and the s-gap is too
small to allow us to attribute the peaks at large voltage to
sub-gap structures. The second group of scattering processes that
leads to above gap series involves Tomasch and McMillan-Rowell
oscillations.\cite{Tomasch,MMR} The Tomasch oscillations are due
to resonances in the superconducting electrode, and their peak
energies are given by

\begin{equation}\label{1}
eV_{n}=\sqrt{(2\Delta)^{2}+[\frac{nhv_{FS}}{2d_{S}}]^{2}},
\end{equation}

\noindent where $\Delta$ is the superconducting gap, $v_{FS}$ is
the Fermi velocity in the electrodes, $d_{S}$ is the
superconducting electrode thickness, and n is the serial number of
the peak. Thus the resonances are not equally spaced, but for
small energy gap values the deviation from a constant voltage
difference between adjacent peaks is quite small, and generally
cannot be observed due to experimental error. McMillan-Rowell
oscillations (MRO),\cite{MMR} are also seen as series of
equidistant peaks in the dynamic resistance spectra above the gap,
and are caused by geometrical resonances of quasiparticles in the
barrier. The voltage difference between adjacent peaks in this
series is given by

\begin{equation}\label{1}
\Delta V=\frac{hv_{FN}}{4ed_{N}},
\end{equation}

\noindent where $v_{FN}$ is the Fermi velocity in the barrier, and
$d_{N}$ is the barrier thickness. We thus find that both the
Tomasch like and McMillan-Rowell like oscillations can yield the
linear behavior of the peak voltage versus peak number seen in
Fig. 5, provided the node gap is small compared to the adjacent
peak spacing. The voltage difference between adjacent peaks
however, depends on either the thickness of the superconducting
electrode in the Tomasch scenario as seen in Eq. (1), or on the
barrier thickness in the MRO case as seen in Eq. (2). Since the
thickness of the superconducting base and cover electrodes is the
same for all our junctions, and the observed series of peaks
depend on the thickness of the barrier as seen in Fig. 5, it seems
that these series are due to McMillan-Rowell like oscillations.
Compared to the original study of MRO,\cite{MMR} the presently
observed conductance peaks are much sharper, especially those in
Fig. 4. As mentioned above, the linear fits in Fig. 5 show that
the ratio of adjacent peaks spacing is approximately equal to the
ratio of the inverse thickness of the barriers. It thus follows
that scaling with the barrier thickness $d_N$ as depicted by Eq.
(2), is found here, and the observed resonances are due to
McMillan-Rowell like oscillations. We stress that this result is
independent of either the barrier strength or the different oxygen
content of the junctions. For a normal metal barrier, Eq. (2)
would have allow us to determine also the Fermi velocity of
quasiparticles in the barrier. In the present case of S/MI/S
junctions with SNS like behavior, analysis yields an effective
velocity $v_{FN} = 1.2\pm0.2 \times 10^{7}$cm/sec, which compares
well with a previous result of $1.5\times 10^{7}$ cm/sec measured
in the same kind of junction with a Fe-doped YBCO barrier. The
later has a much lower resistivity value at low temperatures, of
the order of $10-20 m\Omega cm$, thus being much closer to a
normal metal than the present Ga doped YBCO. A-priori the Fermi
velocity is not well defined here since there is no Fermi surface
at all in isolated VRH materials. When the thin VRH layer however,
is in contact with a superconductor like in the present junctions,
it is possible that the Fermi surface is recovered, and the Fermi
velocity is thus well defined. The fact that similar numbers were
obtained for $v_{FN}$ of the Fe and Ga doped YBCO barriers which
have very different resistivity values, further supports the
notion that the S/MI/S junctions have N like features. These
$v_{FN}$ values of course are only approximate ones. They are
quite smaller than the value $v_{F}\approx 2.5\times 10^7$
obtained by ARPES along the node direction in differently doped
LSCO crystals.\cite{Shen} This ARPES study also shows that there
is no direct link between the measured Fermi velocity in the
cuprates and the size of the resistivity.\\

Since the McMillan-Rowell like oscillations occur at energies
above the gap, the voltage of the first peak in each series
constitutes an upper limit on the magnitude of the gap energy.
Previously, Nesher and Koren measured McMillan-Rowell oscillations
in junctions that were directed along a main crystallographic axis
of YBCO where the d-gap is at it's
maximum.\cite{NesherSubHarmonic} The first peak in their series
appears at 16 mV, which is an upper limit on the value of the
dominant d-wave component of the gap in the 55K phase of
underdoped YBCO. In the present study, on the contrary, the
junctions are aligned along the node direction where the dominant
d-wave gap vanishes, and we can thus measure an upper limit on the
gap energy of the sub-dominant component. The first peak (knee) in
Fig. 3 is found at 3.4 mV, and its voltage in Fig. 4 is 5-6 mV.
This yields an upper limit on the value of the s-gap in the range
of 3.4-6 meV, which is in reasonable agreement with the measured
gap values of 2.5$\pm$0.5 meV found
previously.\cite{Koren2002,KorenJLTP} We note however, that if we
take the average value of the intercepts of the four straight
lines in Fig. 5, we obtain a value of 3.7 mV. This represents a
four series average of the voltage of the first peak, which is
thus a more reliable upper limit on the energy of the s-gap
near the interface.\\

 \maketitle{\large CONCLUSIONS}\\

The present study shows that the experimental properties of S/MI/S
junctions made of underdoped YBCO have several common features
with normal SNS junctions. Measurements of dynamic resistance
spectra in underdoped YBCO junctions along the node direction,
show geometrical resonances in the barrier which behave like
McMillan-Rowell oscillations above the gap. From these we find an
upper limit of 3.7 meV on the magnitude of the
$is$ component of the gap near the interface.\\

{\em Acknowledgments:}  This research was supported in part by the
Israel Science Foundation, the Heinrich Hertz Minerva Center for
HTSC, the Karl Stoll Chair in advanced materials, and by the Fund
for the Promotion of Research at the Technion.\\

\bibliography{AndDepBib.bib}

\bibliography{apssamp}

\end{document}